\documentclass[sigconf,authorversion,natbib=true]{acmart}
\AtBeginDocument{%
  }
\copyrightyear{2026}
\acmYear{2026}
\setcopyright{cc}
\setcctype{by}
\acmConference[SIGIR '26]{Proceedings of the 49th International ACM SIGIR Conference on Research and Development in Information Retrieval}{July 20--24, 2026}{Melbourne, VIC, Australia}
\acmBooktitle{Proceedings of the 49th International ACM SIGIR Conference on Research and Development in Information Retrieval (SIGIR '26), July 20--24, 2026, Melbourne, VIC, Australia}
\acmDOI{10.1145/3805712.3809975}
\acmISBN{979-8-4007-2599-9/2026/07}

\usepackage{bbm}
\usepackage{multirow}
\usepackage{enumitem}
\usepackage{tabularx}
\usepackage{array}
\usepackage{setspace}
\usepackage{balance} 

\newcolumntype{L}{>{\setstretch{0.6}\hangafter=1\hangindent=0.25em}X}

\usepackage[normalem]{ulem}
\useunder{\uline}{\ul}{}

\begin{document}

\title{Sparse Contrastive Learning for Content-Based Cold Item Recommendation}

\author{Gregor Meehan}
\email{gregor.meehan@qmul.ac.uk}
\orcid{0009-0007-2619-9299}
\affiliation{%
  \institution{Queen Mary University of London}
  \city{London}
  \country{United Kingdom}
}

\author{Johan Pauwels}
\email{j.pauwels@qmul.ac.uk}
\orcid{0000-0002-5805-7144}
\affiliation{%
  \institution{Queen Mary University of London}
  \city{London}
  \country{United Kingdom}
}

\renewcommand{\shortauthors}{Meehan and Pauwels}

\begin{abstract}
Item cold-start is a pervasive challenge for collaborative filtering (CF) recommender systems. Existing methods often train cold-start models by mapping auxiliary item content, such as images or text descriptions, into the embedding space of a CF model. However, such approaches can be limited by the fundamental information gap between CF signals and content features. In this work, we propose to avoid this limitation with purely content-based modeling of cold items, i.e.\ without alignment with CF user or item embeddings. We instead frame cold-start prediction in terms of item-item similarity, training a content encoder to project into a latent space where similarity correlates with user preferences. We define our training objective as a sparse generalization of sampled softmax loss with the $\alpha$-entmax family of activation functions, which allows for sharper estimation of item relevance by zeroing gradients for uninformative negatives. We then describe how this \textbf{S}ampled \textbf{E}nt\textbf{m}ax for \textbf{Co}ld-start (\textbf{SEMCo}) training regime can be extended via knowledge distillation, and show that it outperforms existing cold-start methods and standard sampled softmax in ranking accuracy. We also discuss the advantages of purely content-based modeling, particularly in terms of equity of item outcomes.

\end{abstract}




\maketitle
\section{Introduction} \label{sec:intro}
Collaborative filtering (CF) recommendation techniques struggle to predict user preferences for unseen items; this cold-start problem~\cite{schein2002methods} is a key challenge in modern recommender systems (RSs). A common solution is to use item content features, such as images or text, to predict user preferences for new items by mapping into the item embedding space of a CF model~\cite{van2013deep, chen2022generative, bai2023gorec, CLCRec, volkovs2017dropoutnet}. However, these methods rely on the alignment of content and CF information, which may be fundamentally different: CF embeddings capture behavioral patterns, while content features capture semantic attributes. Recent work~\cite{meehan2025inherited} shows that this mismatch can lead to unreliable predictive behavior, as cold-start models attempt to predict item popularity based on item content alone. 

In this work, we propose to avoid these limitations by modeling cold-start prediction in terms of item-item similarities in the content space. Instead of attempting to match CF and content signals, we train a content encoder to project into a latent space where item similarity is correlated with user preference. This objective naturally leads to a contrastive training regime~\cite{SimCLR,oord2018representation} with supervision from user-item interactions, making sampled softmax loss~\cite{wu2024effectiveness} an appropriate choice. However, softmax (pseudo-)probability outputs are dense, i.e.\ all non-zero; for large contrastive batches, this may dilute the supervision signal with gradient updates on largely irrelevant negative examples. This is particularly significant for content-based RS modeling, where item representations are less flexible due to their dependence on multimodal content features, potentially introducing noise into item-item similarities~\cite{shang2024improving}. 

We therefore generalize sampled softmax using the $\alpha$-entmax~\cite{peters2019sparse} family of Fenchel-Young losses~\cite{blondel2020learning}, which interpolate between softmax and sparsemax~\cite{martins2016softmax} for controllable sparsity in the output probabilities. We show that this \textbf{S}ampled \textbf{E}nt\textbf{m}ax for \textbf{Co}ld-start (\textbf{SEMCo}) framework\footnote{To support reproducibility of our work, we release our code publicly at \href{https://github.com/gmeehan96/SEMCo}{https://github.com/gmeehan96/SEMCo}.} admits a natural form of knowledge distillation~\cite{hinton2015distilling} in the content space, transferring item-item similarity structure from a higher-capacity teacher encoder to a lightweight student. Through experiments on four multimodal datasets, we show that SEMCo significantly outperforms cold-start baselines in cold item ranking accuracy. Furthermore, by decoupling training from the popularity-biased supervision of CF models, SEMCo naturally produces more equitable outcomes across the item population without sacrificing overall accuracy or requiring explicit fairness regularization. We also find that the refined gradient signal allows sparse $\alpha$-entmax supervision to outperform standard sampled softmax, with particular benefits for knowledge distillation.

\section{Related Work}
\paragraph{\textbf{Content-Based Cold-Start}} Item content (e.g.\ images, text descriptions, audio) is often leveraged to solve the cold-start problem~\cite{zhang2025cold}. Dropout-based methods, including the classic DropoutNet~\cite{volkovs2017dropoutnet} and its more recent extensions~\cite{zhu_recommendation_2020, CLCRec,zhou2023contrastive,wang2024preference}, randomly swap between CF and content embeddings during training to simulate cold-start. Other `generative' methods~\cite{van2013deep,bai2023gorec,huang_aligning_2023,chen2022generative,sun2020lara,tamm2026leveraging} train a content encoder to generate synthetic cold item CF embeddings by mapping into the embedding space of a pre-trained CF model.

Recent work~\cite{meehan2025inherited} suggests that the information gap between item content and CF signals is a key challenge for these content-based solutions. In particular, generative cold-start models exhibit erratic predictive behavior due to their imitation of popularity bias~\cite{klimashevskaia2024survey} in the supervisory CF model. We show below that SEMCo avoids this behavior by abandoning CF model supervision.

\paragraph{\textbf{RS Knowledge Distillation}}
RS knowledge distillation~\cite{hinton2015distilling} (KD) typically treats user-item scores~\cite{huang_aligning_2023, lee2019collaborative, tang2018ranking,zhuang2025bridging} or ranking lists~\cite{zhu2020ensembled, kang2023distillation, kang2025unbiased} as the knowledge to be passed from a large teacher RS to a lightweight student RS. Generative cold-start methods~\cite{bai2023gorec, huang_aligning_2023, chen2022generative} are a form of KD where the teacher is a pre-trained CF model, and the student is a content-based RS. Our SEMCo distillation variants instead distill in the content space, passing knowledge from a larger to smaller content-based RS and avoiding the information gap between content and CF representations.

\paragraph{\textbf{RSs and $\alpha$-Entmax}} Entmax and sparsemax have previously been applied for sparse multi-head attention in session-based RSs~\cite{choi2024multi, 10191924, yuan2021dual, maneeroj2022end}. While some works~\cite{pu2024learning, kong2020rankmax} explore entmax-related RS losses, to our knowledge these have not been applied in content-based or cold-start RS contexts, nor for RS KD training.

\section{Sampled Entmax for Cold-Start}\label{sec:method}
\subsection{Preliminaries}
\subsubsection{Problem Definition} We write $\mathcal{U}$ and $\mathcal{I}$ for our user and item sets respectively, with $\mathbf{R}\in\{0,1\}^{|\mathcal{U}|\times |\mathcal{I}|}$ our binary interaction matrix. Vectors are written in lowercase (e.g. $\mathbf{x}$), and matrices uppercase (e.g. $\mathbf{X}$); 
each item $i$ has content feature vectors $\mathbf{x}_1,...,\mathbf{x}_M$ in $M$ data modes. Following previous cold-start works~\cite{huang_aligning_2023,chen2022generative,CLCRec}, we assume all modes are present during training and inference, and focus on feature vectors generated by pre-trained models.

\subsubsection{Sparsemax and $\alpha$-Entmax}\label{sec:entmax}

The $\alpha$-entmax activation~\cite{peters2019sparse} generalizes softmax 
to efficiently compute sparse probabilities.  Given $n$-dimensional input logits $\mathbf{z}$, it solves the maximization problem
\begin{equation}
\alpha\text{-entmax}(\mathbf{z})
\;=\;
\operatorname*{argmax}_{\mathbf{p} \in \Delta_d}
\;\; \mathbf{p}^\top \mathbf{z} + H_\alpha(\mathbf{p})
\end{equation}
for the probability simplex $\Delta_n$ and Tsallis $\alpha$-entropies $H_\alpha(\mathbf{p})$~\cite{tsallis1988possible}, with $1\leq\alpha\leq2$. When $\alpha=1$, this is equivalent to softmax; otherwise, it is solved by finding a unique threshold $\eta$:
\begin{equation}
\alpha\text{-entmax}(\mathbf{z})
=
\bigl[(\alpha - 1) \mathbf{z} - \eta \mathbf{1}\bigr]_+^{\frac{1}{\alpha - 1}},
\end{equation}
where $\mathbf{1}$ is the ones-vector and $\left[\cdot\right]_+$ is the positive-part operator. When $\alpha=2$, this is the \textbf{sparsemax} function~\cite{martins2016softmax}. 

Alongside this activation, \cite{peters2019sparse} also defines a Fenchel-Young~\cite{blondel2020learning} extension of the standard cross-entropy loss for $\alpha$-entmax~\cite{peters2019sparse}. For target probabilities $\mathbf{p}$ and predicted probabilities $\mathbf{\hat{p}}=\alpha\text{-entmax}(\mathbf{z}/\tau)$ on logits $\mathbf{z}$ with temperature hyperparameter $\tau$, this is 
\begin{equation}\label{eq:loss}
\mathcal{L}_{\alpha,\tau} (\mathbf{z},\mathbf{p})
=
\bigl(\mathbf{\hat{p}} - \mathbf{p}\bigr)^\top \mathbf{z} + H_\alpha(\mathbf{\hat{p}}).
\end{equation}
We modify the original formulation to include $\tau$ for alignment with RS sampled softmax~\cite{wu2024effectiveness}, as discussed below in Section \ref{sec:training}.   

Notably, since $\mathcal{L}_{\alpha,\tau}$ is a Fenchel-Young loss, its gradient is 
\begin{equation}
    \nabla\mathcal{L}_\mathbf{z}(\mathbf{z},\mathbf{p}) =  \alpha\text{-entmax}(\mathbf{z})- \mathbf{p},
\end{equation} 
so when $\alpha>1$ there are no gradient updates from logits less than the threshold $\eta/(\alpha-1)$. In an RS, this means that model updates are focused on the most relevant items, whereas all gradients are non-zero in dense softmax probabilities, including those for uninformative negatives. We show that this gradient sparsity can facilitate improved content similarity modeling for cold-start RSs.

\subsection{Content-Based Modeling}
\subsubsection{User Preference Prediction} To avoid the pitfalls of directly aligning content features with CF embeddings, we model user preferences using item content alone. Similarly to CF linear autoencoders (LAEs)~\cite{ning2011slim, steck2019embarrassingly, vanvcura2022scalable, vanvcura2025evaluating, park2025normalization}, we calculate preferences via the matrix multiplication $\mathbf{R}\mathbf{B}$, where $\mathbf{B}\in \mathbb{R}^{|\mathcal{I}|\times |\mathcal{I}|}$ captures item-item relationships. For warm items, LAEs learn $\mathbf{B}$ directly from user-item interactions; however, `appending' columns to $\mathbf{B}$ for new items would require querying the entire warm item set, presenting scalability and latency challenges. We therefore follow shallow LAEs~\cite{vanvcura2022scalable,vanvcura2025evaluating,vanvcura2024beeformer} in factorizing $\mathbf{B}$ so that $\mathbf{B} = \mathbf{Y}\mathbf{Y}^\top$, where 
\begin{equation}
    \mathbf{Y}=f(\mathbf{X}_1,...,\mathbf{X}_M)\in\mathbb{R}^{|\mathcal{I}|\times d}
\end{equation}
is a $d$-dimensional encoding of the item content features, with each row L2-normalized. Then, we write our preference scores as 
\begin{equation}
    \mathbf{R}\mathbf{B} = \mathbf{R}(\mathbf{Y}\mathbf{Y}^\top) =  (\mathbf{RY})\mathbf{Y}^\top,
\end{equation}
and new item preferences can be predicted by dot products between their encoded content and the rows of user embedding matrix $\mathbf{RY}$. 

\subsubsection{Content Encoder}\label{sec:content_encoder} Our content encoder $f(\cdot)$ first maps each content input $\mathbf{x}_m$ with a single layer followed by batch normalization and ReLU. We fuse the resulting hidden vectors $\mathbf{h}_m$ via a simple attention module: we concatenate to form $\mathbf{h} = [\mathbf{h_1};...;\mathbf{h}_M]$ and project with a two-layer MLP with hidden ReLU and softmax output to form attention weights $\mathbf{a}\in\mathbb{R}^M$. The final hidden output is $\tilde{\mathbf{h}} = \sum_{m=1}^M a_i \mathbf{h}_m$; we transform $\tilde{\mathbf{h}}$ with a linear layer, then L2-normalize to produce our final encoded content output $\mathbf{y}$.

\subsection{Training}\label{sec:training}
\subsubsection{Sampled Entmax}
For cold item recommendation, we want similarity in our encoded content vectors to correlate with content-based user preferences. We therefore choose a contrastive approach~\cite{oord2018representation,SimCLR} based on the common sampled softmax RS loss~\cite{wu2024effectiveness} with in-batch negative sampling. Batches are constructed with positive user-item pairs $\mathcal{B}\subset\{(u,i): \mathbf{R}_{ui}=1\}$, and loss is calculated as a per-user classification task over all items in the batch. 

We can define a generalized form of sampled softmax for $\alpha$-entmax activations. For user $u$ at batch index $j$, the classification target is the indicator vector $\mathbf{e}_j$ at index $j$, while the logit vector $\mathbf{z}_u$ is the cosine similarity between the user vector and each batch item. The corresponding \emph{sampled entmax loss} for $u$ is 
\begin{equation}\label{eqn:sem_loss}
    \mathcal{L}^\textrm{SEM}(u) =\mathcal{L}_{\alpha,\tau} (\mathbf{z_u}/\tau_0,\mathbf{e}_j).
\end{equation}
We find that including the additional scaling factor $\tau_0$ on the logits improves training stability for $\alpha>1$, with $\tau_0=\tau$. In the softmax case ($\alpha=1$), we set $\tau_0=1$, so Equation \ref{eqn:sem_loss} is equivalent to standard sampled softmax. This general form allows us to test the effectiveness of other $\alpha$ values for training our content-based cold-start model. In particular, we are interested in whether increased sparsity in the predicted user-item probabilities can improve training quality by prioritizing gradient updates for the most relevant items.

\subsubsection{Item-Item Similarity Distillation}\label{sec:distil}
SEM is a natural fit for the well-explored application of KD in classification contexts ~\cite{hinton2015distilling,gou2021knowledge}. We choose to apply KD to item-item similarities (as these are `atomic' in SEMCo) by distilling from larger-dimension content embeddings to smaller ones. Given a teacher-encoded content matrix $\mathbf{Y}_T\in \mathbb{R}^{d_T}$, we project $\mathbf{Y}_T$ to a smaller dimension $d$ with a two-layer MLP with hidden ReLU activation. Then, for each item $c$ in a batch item set $\mathcal{C}$, we calculate teacher ($\mathbf{z}_{c,T}$) and student ($\mathbf{z}_{c,S}$) cosine similarities to all other items in $\mathcal{C}$, and define our distillation loss as 
\begin{equation}\label{eqn:distil_loss}
    \mathcal{L}^\textrm{Distill}(c) =\mathcal{L}_{\alpha,\omega} \left(\frac{\mathbf{z_{c,S}}}{\omega_0},\alpha\text{-entmax}\left(\frac{\mathbf{z_{c,T}}}{\omega_0}\right)\right)
\end{equation}
for distillation temperature $\omega$. Rather than using the items in batch $\mathcal{B}$ to form $\mathcal{C}$, we sub-sample the users from $\mathcal{B}$ and choose several positive examples per user; the resulting intra-user pairs are more likely to provide informative supervision to the student. We again aim to leverage the sparsity induced by the $\alpha$-entmax transformation of the teacher logits, so that the distillation signal is focused on the item-item similarities deemed most relevant by the teacher.

We combine $\mathcal{L}^\textrm{Distill}$ with the SEM loss for the final objective 
\begin{equation}
    \mathcal{L}^\text{Total} = \mathcal{L}^\textrm{Distill} + \lambda \mathcal{L}^\textrm{SEM}
\end{equation}
with hyperparameter $\lambda$ controlling the influence of the sampled entmax loss. We test both offline and online distillation, i.e.\ pre-training the teacher or training it alongside the student. In the online case, we supervise the teacher with $\mathcal{L}^\textrm{SEM}$ and use an exponential moving average~\cite{tarvainen2017mean} of the teacher's outputs as the student's input, similarly to `early-exit' self-distillation~\cite{zhang2019your,zhang2021self}.

\section{Experiments}
\subsection{Experimental Design}
\subsubsection{Datasets}
We measure cold-start performance on four datasets in E-commerce, music, and micro-video; these are summarized in Table \ref{table:data_stats}. \textbf{Clothing} and \textbf{Electronics} are commonly used subsets of the Amazon E-commerce datasets~\cite{mcauley2015image}: we use the 5-core filtered interaction data and pre-processed image and text features available via the MMRec framework~\cite{zhou2023mmrec}. We exclude the visual features for Clothing, following other works~\cite{zhang_modality-balanced_2024,meehan2025inherited, malitesta2023popularity} which find these can decrease performance. \textbf{Microlens}~\cite{ni2023content} is a micro-video dataset with pre-processed image, text, and video features, also available via MMRec. Music4All-Onion~\cite{Music4All-Onion} (\textbf{M4A-Onion})\footnote{We define our M4A-Onion subset as follows: first, we exclude skipped songs, i.e.\ where the next song starts within 30 seconds, as in~\cite{seshadri2024enhancing}.  Then, as in~\cite{ganhor2024multimodal, meehan2025artist}, we only consider interactions from 2018 and users age 10 to 80, before excluding user-item pairs that only occur once and 5-core filtering the users and items. We generate text content features with BERT~\cite{BERT} on M4A-Onion's pre-processed lyrics and audio features with music representation learning model MusicFM~\cite{won2024foundation} on the 30-second audio clips.} contains multimodal features for the songs in music dataset Music4All~\cite{santana2020music4all} and user-song interactions from Last.fm listening sessions~\cite{schedl2022lfm}.   

\subsubsection{Evaluation}
As in previous studies~\cite{chen2022generative, huang_aligning_2023, meehan2025inherited}, we simulate cold-start by randomly selecting 20\% of items to exclude from training, dividing them 50/50 for validation/test. The remaining interactions are split 80/10/10 into warm training/validation/test sets for the baselines' warm backbone model. To measure cold test set performance, we use user-oriented ranking metrics Recall@$k$ and NDCG@$k$, as well as Item Mean Discounted Gain@$k$ (MDG@$k$)~\cite{zhu2021fairness}, which evaluates item-oriented ranking accuracy~\cite{meehan2025inherited}. We also measure fairness of item exposure in model predictions with Gini diversity~\cite{adomavicius2011improving}. All results are averaged over five runs at optimal hyperparameters. 

\begin{table}[]
    \caption{Dataset statistics. The size of the pre-processed feature vector for each data mode is indicated in parentheses. \label{table:data_stats}}
  \small
  \setlength\extrarowheight{-1.25pt}
  \begin{tabular}{cccccc}
  \toprule
  Dataset                 & Users    & Items & Interactions & Density & Modes (dim)   \\ \midrule
  Clothing   & 39,387    & 23,033 & 278,677 & 0.031\% & Text (384) \\[1.5pt] 
  \multirow{2}{*}{Electronics} & \multirow{2}{*}{192,403}  & \multirow{2}{*}{63,001} & \multirow{2}{*}{1,689,188} & \multirow{2}{*}{0.014\%} & Image (4096) \\ 
  &                          &                         &                            &                          & Text (384) \\[1.5pt] 
     \multirow{2}{*}{M4A-Onion} & \multirow{2}{*}{8,807}  & \multirow{2}{*}{43,886} & \multirow{2}{*}{1,510,034} & \multirow{2}{*}{0.391\%} & Audio (1024) \\ 
  &                          &                         &                            &                          & Text (768) \\[1.5pt]  
  
  \multirow{3}{*}{Microlens} & \multirow{3}{*}{98,129}  & \multirow{3}{*}{17,228} & \multirow{3}{*}{705,174} & \multirow{3}{*}{0.042\%} & Image (1024) \\
  &                          &                         &                            &                          & Text (1024) \\
   &                          &                         &                            &                          &Video (768) 
  \\ \bottomrule
  \end{tabular}

  \end{table}

\subsubsection{Baselines}
We select five content-based methods as our cold-start baselines: \textbf{CLCRec}~\cite{CLCRec} trains from scratch with a contrastive objective to align content and CF vectors, while the other four methods (\textbf{ALDI}~\cite{huang_aligning_2023}, \textbf{GAR}~\cite{chen2022generative}, \textbf{GoRec}~\cite{bai2023gorec}, and \textbf{Heater}~\cite{zhu_recommendation_2020}) train content encoders to map into the latent space of a pre-trained CF model. Following \cite{meehan2025inherited}, we use \textbf{FREEDOM}~\cite{zhou_tale_2023} as this supervisory warm model, and create the content inputs by L2-normalizing then concatenating the single feature vectors required by the cold models. Hyperparameters are tuned by grid search on cold validation NDCG@20 using the ranges described in the original papers.

\begin{table*}[]
\caption{Cold test set Recall@20 (R@20), NDCG@20 (N@20), and MDG@20 (M@20). We {\ul underline} the best model in each group and \textbf{bold} the best model overall. Improv. is calculated over the best-performing cold-start baseline for Sparsemax SEMCo-Online. 
}\label{tab:cold_results}
\small
\centering
\setlength\extrarowheight{-1.25pt}
\begin{tabular}{c@{\hskip6.0pt}c c@{\hskip6.0pt}c@{\hskip6.0pt}c@{\hskip6.0pt}c@{\hskip6.0pt}c      c@{\hskip6.0pt}c@{\hskip6.0pt}c c@{\hskip6.0pt}c@{\hskip6.0pt}c c@{\hskip6.0pt}c c}
\toprule
       &  & \multicolumn{5}{c}{Cold-Start Baselines} &       \multicolumn{3}{c}{SEMCo}     & \multicolumn{3}{c}{SEMCo-Offline}  & \multicolumn{2}{c}{SEMCo-Online} &                     
       \\ \cmidrule(l{3pt}r{3pt}){3-7} \cmidrule(l{3pt}r{3pt}){8-10} \cmidrule(l{3pt}r{3pt}){11-13} \cmidrule(l{3pt}r{4pt}){14-15}
Dataset                     & Metric  & ALDI   & CLCRec         & GAR    & GoRec        & Heater & Soft.  & Ent.         & Sparse.      & Soft.           & Ent.            & Sparse.         & Ent.   & Sparse.          & \textit{Improv.}\\ \midrule
\multirow{4}{*}{Clothing}    & R@20   & 0.0954 & 0.1190         & 0.1148 & {\ul 0.1389} & 0.1225 & 0.1408 & {\ul 0.1435} & 0.1432       & 0.1419          & 0.1398          & {\ul 0.1435}    & 0.1457 & \textbf{0.1486}  &7.0\%      \\
                             & N@20   & 0.0453 & 0.0599         & 0.0541 & {\ul 0.0638} & 0.0561 & 0.0684 & 0.0697       & {\ul 0.0699} & 0.0703          & 0.0702          & {\ul 0.0715}    & 0.0701 & \textbf{0.0717}  &12.4\%   \\
                             & M@20   & 0.0312 & 0.0462         & 0.0372 & {\ul 0.0495} & 0.0398 & 0.0555 & 0.0567       & {\ul 0.0572} & \textbf{0.0584} & 0.0581          & 0.0571          & 0.0562 & {\ul 0.0582}           &17.7\%  \\
                             & Gini   & 0.4415 & {\ul 0.6185} & 0.3516          & 0.4840 & 0.3630 & 0.6316 & 0.6694          & {\ul 0.6841} & 0.6698 &  \textbf{0.7106} & 0.6871          & 0.5989 & {\ul 0.6307} & 2.0\% \\[2pt]
\multirow{4}{*}{Electronics} & R@20   & 0.0297 & 0.0299         & 0.0321 & {\ul 0.0376} & 0.0329 & 0.0462 & 0.0458       & {\ul 0.0481} & 0.0445          & {\ul 0.0493}    & 0.0492          & 0.0514 & \textbf{0.0526}  &39.7\%    \\
                             & N@20   & 0.0130 & 0.0138         & 0.0148 & {\ul 0.0171} & 0.0149 & 0.0210 & 0.0211       & {\ul 0.0217} & 0.0201          & {\ul 0.0223}    & {\ul 0.0223}    & 0.0230 & \textbf{0.0236}  &38.0\%    \\
                             & M@20   & 0.0061 & 0.0090         & 0.0074 & {\ul 0.0092} & 0.0069 & 0.0195 & {\ul 0.0198} & 0.0194       & 0.0188          & 0.0206          & {\ul 0.0207}    & 0.0211 & \textbf{0.0214}  &133.7\%     \\
                             & Gini   & 0.1759 & {\ul 0.2864} & 0.2483          & 0.1523 & 0.1279 & 0.6381 & \textbf{0.6749} & 0.6233          & 0.5578 & 0.6117          & {\ul 0.6265} & 0.5950 & {\ul 0.6201} & 116.5\% \\[2pt]
\multirow{4}{*}{M4A-Onion}   & R@20   & 0.0406 & {\ul 0.0577}   & 0.0464 & 0.0299       & 0.0544 & 0.0777 & {\ul 0.0825} & 0.0806       & 0.0831          & {\ul 0.0886}    & 0.0863          & 0.0889 & \textbf{0.0910}  &57.7\%  \\
                             & N@20   & 0.0403 & {\ul 0.0557}   & 0.0467 & 0.0316       & 0.0557 & 0.0743 & {\ul 0.0792} & 0.0772       & 0.0783          & 0.0851          & {\ul 0.0852}    & 0.0854 & \textbf{0.0887}  &59.3\%    \\ 
                             & M@20   & 0.0114 & {\ul 0.0159}   & 0.0128 & 0.0104       & 0.0152 & 0.0237 & {\ul 0.0251} & 0.0246       & 0.0249          & \textbf{0.0268} & \textbf{0.0268} & 0.0259 & {\ul 0.0267}           &67.9\%     \\
                             & Gini   & 0.1867 & 0.1862          & {\ul 0.2014} & 0.1765 & 0.1667 & 0.4638 & 0.4915          & {\ul 0.4939} & 0.4739 & \textbf{0.5280} & 0.4879          & 0.4677 & {\ul 0.5010} & 148.7\% \\[2pt]
\multirow{4}{*}{Microlens}   & R@20   & 0.1075 & 0.0914         & 0.1180 & {\ul 0.1294} & 0.1203 & 0.1336 & {\ul 0.1372} & 0.1332       & 0.1391          & {\ul 0.1411}    & 0.1371          & 0.1387 & \textbf{0.1455}  &12.5\%      \\
                             & N@20   & 0.0445 & 0.0401         & 0.0487 & {\ul 0.0562} & 0.0504 & 0.0590 & {\ul 0.0600} & 0.0594       & {\ul 0.0619}    & 0.0615          & 0.0607          & 0.0615 & \textbf{0.0629}  &11.9\%           \\
                             & M@20   & 0.0328 & 0.0325         & 0.0359 & {\ul 0.0448} & 0.0384 & 0.0504 & {\ul 0.0511} & 0.0488       & 0.0532          & 0.0522          & \textbf{0.0537} & 0.0499 & {\ul 0.0524}           &16.8\%     \\
                             & Gini   & 0.3882 & 0.3760          & {\ul 0.4041} & 0.3167 & 0.3267 & 0.5409 & {\ul 0.6155} & 0.6052          & 0.5182 & 0.5457          & {\ul 0.6238} & 0.5054 & {\ul 0.5599} & 38.5\%  \\
 \bottomrule        
\end{tabular}
\end{table*}

\subsubsection{Implementation Details} We test three variants of our method: \textbf{SEMCo} alone and the offline and online distillation approaches described in Section \ref{sec:distil} (\textbf{SEMCo-Offline} and \textbf{SEMCo-Online}). We evaluate the softmax ($\alpha=1$), 1.5-entmax, and sparsemax ($\alpha=2$) versions of our SEMCo losses, as there are efficient algorithms at these $\alpha$ values for selecting $\eta$ in at most $\mathcal{O}(d \log d)$ time \cite{peters2019sparse}.

All models, including baselines, use the Adam optimizer~\cite{Kingma2014AdamAM} and have output dimension 64. For SEMCo, the batch size is 2048 and the hidden size is 192; we decay the learning rate with a cosine schedule~\cite{loshchilov2016sgdr} from 0.001 to zero over 15 epochs, and perform a grid search on the temperature $\tau$ and L2 regularization coefficient. Full parameter search ranges for all methods are on our GitHub.  The teacher hidden and output sizes are 384 for both distillation methods. For SEMCo-Offline, we conduct a grid search on the teacher settings, then a random search on the student model's temperature, distillation temperature, number of positive examples, and SEM weight $\lambda$. Due to the added complexity of jointly tuning the student and teacher in SEMCo-Online, we use TPE~\cite{bergstra2011algorithms} for more efficient search. We linearly warm-up the online student's learning rate from 0 to 0.001.

\subsection{Results}
Table \ref{tab:cold_results} displays cold-start results for our SEMCo variants and baselines. We observe that SEMCo consistently outperforms the cold-start baselines; with the exception of Softmax SEMCo and Entmax SEMCo-Offline in Clothing R@20, the gains in ranking metrics of all SEMCo methods over the best-performing baseline are statistically significant in all datasets, as measured by paired t-tests ($p<0.05$). This illustrates the effectiveness of content similarity-based modeling for cold item prediction, and suggests the alignment of content with CF embeddings can limit existing methods.

\begin{figure}[]
\begin{tabular}{c}
     \includegraphics[width=60mm,trim={0.3cm 0.3cm 0.3cm 0}]{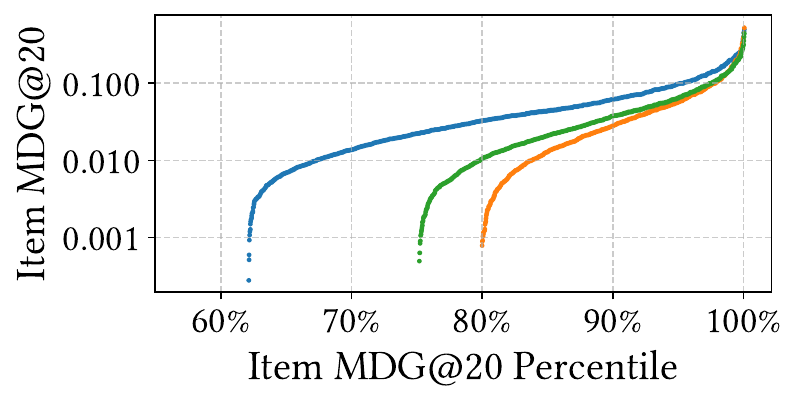} \\ 
     \includegraphics[width=60mm,trim={0.3cm 0.3cm 0.3cm 0}]{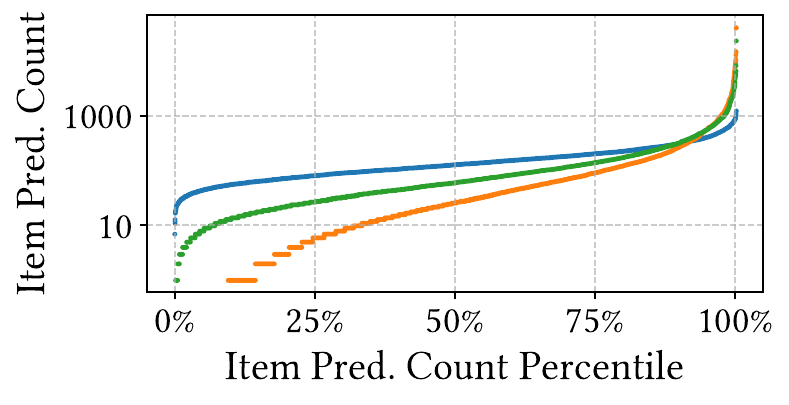} 
\end{tabular}
\includegraphics[width=70mm,trim={0 0.7cm 0 0.0cm}]{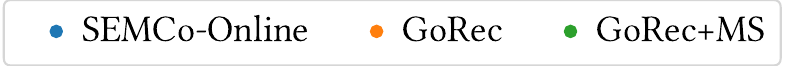}
\caption{Item MDG@20 and prediction counts across the cold item population in the Electronics dataset. Each dot represents an item, and items with zero MDG or prediction count are not pictured. MS stands for magnitude scaling~\cite{meehan2025inherited}.} \label{fig:elec_pop}

\end{figure}

The notable gains in MDG@20 (up to 133.7\% on Electronics) and Gini diversity (up to 148.7\% on M4A-Onion) also show that SEMCo produces more equitable item outcomes than previous methods. On the Clothing dataset, CLCRec has comparable diversity to SEMCo, but much lower ranking metrics, i.e.\ SEMCo provides a better balance between fairness and overall accuracy. We visualize this outcome equity further in Figure \ref{fig:elec_pop} across the Electronics item population, comparing with the best-performing cold-start baseline (GoRec); we also include post-training magnitude scaling (GoRec+MS) bias mitigation as proposed in \cite{meehan2025inherited}. We observe first that all methods have no successful recommendations for at least 60\% of the item population. However, by avoiding CF biases, SEMCo-Online's item accuracies are more balanced than GoRec's, even with additional bias mitigation applied. 

The source of the gains in Gini diversity is also clear, as SEMCo-Online's item prediction counts are much more evenly distributed across the population, with nearly all items predicted at least 10 times. This contrasts with GoRec, where there is a sharp peak for a small number of items, while 10\% of items are never recommended at all, violating the Max-Min fairness principle for judging outcome fairness in RSs~\cite{zhu2021fairness}. Unlike many methods in bias mitigation, which produce more balanced outcomes at the cost of overall accuracy~\cite{klimashevskaia2024survey}, we see that SEMCo's content-based prediction can increase predictive fairness while also improving accuracy from the user perspective.

\subsubsection{SEMCo Variants} We now analyze the impact of different $\alpha$ values and training strategies in SEMCo. Either Entmax or Sparsemax performs best with the base SEMCo loss across all datasets, illustrating the value of introducing sparsity into the standard sampled softmax-style training in content-based contexts. Although the cost of computing these activation functions ($\mathcal{O}(d \log d)$) is slightly higher than for softmax ($\mathcal{O}(d)$),  we find in practice that this only increases training time by 10-20\%. We see that Entmax and Sparsemax also dominate the offline distillation group, which in most cases provides a slight improvement over SEMCo.  

Due to this superior performance of Entmax and Sparsemax in the base and offline settings, we focus on these variants for SEMCo-Online.  Online distillation benefits from the student evolving alongside the teacher, providing a form of curriculum supervision~\cite{zeng2022curriculum}.  We see that Sparsemax profits more from the online method than Entmax, with Sparsemax Online producing the best user ranking accuracy in all datasets, up to a 14.9\% gain over the base SEMCo Sparsemax in M4A-Onion N@20. As discussed in Section \ref{sec:method}, we hypothesize that the sparser supervision signals from the Sparsemax teacher allow the student to learn the most important item-item similarities more effectively; this is supported by observing that, during training, typically less than 10\% of entries are non-zero in the Sparsemax student and teacher probability matrices used in Equation \ref{eqn:distil_loss}. By contrast, over 70\% of entries are non-zero for Entmax. As with any sparse modeling, there is a crucial balance in increasing sparsity to emphasize key supervision signals without removing them entirely; our results suggest that Sparsemax manages this balance more effectively in the online distillation setting.

Although only requiring a single training process, online distillation presents challenges in training stability and joint tuning of student and teacher hyperparameters~\cite{moslemi2024survey}. The optimal choice between SEMCo-Online and -Offline  may therefore depend on dataset attributes and training cost, and the base SEMCo provides a simpler alternative with slightly lower accuracy. We also note that, as described, SEMCo constructs user embeddings by a matrix multiplication across the full item set, raising scalability concerns for large item catalogs; however, scalability  can be improved by only considering `active' batch items during training and inference~\cite{vanvcura2025evaluating}.

\section{Conclusion}
In this paper, we present our SEMCo method for training a content-based model to address item cold-start, building on the intuition that content-based user preferences can be captured by item-item similarities. We propose a contrastive training regime that generalizes sampled softmax loss to induce sparsity in the predicted probabilities during training with $\alpha$-entmax, and describe how this method can be extended with offline or online knowledge distillation to further refine the learned item-item similarities. The resulting models significantly outperform previous cold-start models in four datasets across E-commerce, micro-video and music, both in terms of user-oriented ranking accuracy and prediction quality and diversity across items. Our entmax-based training objective facilitates further performance gains, particularly with Sparsemax in SEMCo-Online. Future work will explore how this method can be extended to missing modality scenarios~\cite{ganhor2024multimodal, malitesta2026training}, as well as how sparsity can be further leveraged to improve the alignment between content similarity and user preference.

%

\begin{acks}
This work was funded by UKRI as part of the UKRI CDT in Artificial Intelligence and Music [grant number EP/S022694/1].
\end{acks}

\bibliographystyle{ACM-Reference-Format}
\balance
\bibliography{bibfile}


\end{document}